\documentclass[prl,twocolumn,showpacs,superscriptaddress]{revtex4}
\usepackage{graphicx}
\usepackage{dcolumn}
\usepackage{bm}

\begin{document}
\title{Solid-solid volume collapse transitions are zeroth order}

\author{S. Bustingorry}
\author{E. A. Jagla}
\affiliation{Consejo Nacional de
Investigaciones Cient\'{\i}ficas y T\'ecnicas, Centro At\'omico Bariloche,
8400 Bariloche, Argentina}
\author{J. Lorenzana}
\affiliation{Center for Statistical 
Mechanics and Complexity, INFM, Dipartimento di Fisica,  
Universit\`a di Roma La Sapienza, P. Aldo Moro 2, 00185 Roma, Italy}
\date{\today}
\begin{abstract}
We present an exactly solvable non-linear elastic
model of a volume collapse transition in an isotropic solid.
Integrity of the lattice  through the transition leads to an
infinite-range density-density interaction, which drives
classical critical behavior.
Nucleation is forbidden within a pressure window leading to 
intrinsic hysteresis and an unavoidable discontinuity 
of the thermodynamic potential (zeroth order transition).
 The window shrinks with increasing temperature ending 
at a critical point at a temperature related 
to the shear modulus. Mixed phases behave non-extensively
and show negative compressibility.
We discuss the implications for Ce, SmS, and related systems.   
\end{abstract}
\maketitle

Room temperature $\gamma$ Cerium shows a well known transition to 
isostructural $\alpha$ Cerium at 
$\sim$ 0.8GPa  with a 17\% reduction of the volume.
 Similar (though not necessarily isostructural) transitions occur in 
other $f$-electron systems\cite{gsc79,law81}. 
Volume-collapse transitions  were also predicted to occur in 
colloidal systems\cite{bol94} and close to 
first order electronic transitions when mesoscopic inhomogeneous electronic 
states are frustrated by the
long-range Coulomb interaction\cite{lor01Ilor02}.

Consider one of these systems which is driven across the 
volume-collapse transition by adiabatically changing the average volume.
In analogy with the van der Waals  liquid-gas transition
one expects a range of average volume values for which
there is phase coexistence at a constant equilibrium
pressure. This presents an interesting nucleation problem. 
Within purely energetic considerations the mixed 
phase consists of a region of the sample with the small unit 
cell volume and a region with the large unit cell volume 
separated by a flat surface so as to minimize the interface boundary energy. 
The lattice mismatch at the interface can be seen as a line of
dislocations. However dislocations are very difficult to nucleate
in a solid since they require the breaking of chemical bonds. Indeed it is
customary  to assume on solid-solid transitions  
that the lattice integrity is preserved\cite{bar84,kar95ras01,kle02}. 
Under this hypothesis the thermodynamic ground state is frustrated
from the onset, and the question arises of how the system minimizes 
its free energy, preserving the integrity constraints.
 To answer this question we solve a non-linear elastic model 
of a volume-collapse transition in an isotropic solid assuming 
lattice integrity (no dislocations). 
The latter condition leads to the well 
known St. Venant compatibility constraints\cite{cha94} among the different 
components of the strain tensor and allows the free energy to
 be expressed as a function of the order parameter (OP) alone, 
with long-range interactions\cite{bar84,kar95ras01,kle02}.
This kind of approach has been shown to lead to anomalous nucleation
in volume conserving transitions\cite{kle02}.

We find that compatibility constraints induce a
distance-independent density-density interaction
and thus mean-field (MF) theory becomes exact and the model  
results to be trivially solvable at all temperatures in the absence
of short-range interactions. Critical exponents are MF like 
even in the presence of short-range interactions. In the mixed 
phase the system behaves non extensively
and violates usual stability criteria for extensive systems. In
particular, it possesses negative compressibility. For an experiment 
in which the pressure acts as the control variable we show 
that usual nucleation is forbidden within a finite pressure
window. This pressure window implies an intrinsic hysteresis loop 
in a pressure controlled experiment.
 Outside the pressure window the transition can 
occur, but only irreversibly with an intrinsic jump of the
thermodynamic potential. Since the discontinuity occurs in the 
thermodynamic potential itself rather than in its first derivative
the transition can be called of zeroth order (instead of first) as 
 done with similar phenomena in gravitational systems\cite{veg02}.
The extent of the hysteresis loop reduces as a function of temperature,
vanishing at a critical point. 
The temperature of the critical point is related to the shear modulus of
 the material.

\paragraph{Model}

The OP is taken to be the dilation strain 
$e_1\equiv\sum_{i=1}^d \epsilon_{ii}/\sqrt{d}$
 where $\epsilon_{ij}$ are the components of the infinitesimal
strain tensor\cite{cha94} $\epsilon$ and $d$ is the 
dimensionality of the system.
Our starting point is a coarse grained\cite{cha95} 
elastic Hamiltonian $H\equiv H_1+H_h$ with  $H_1$
a non-linear Ginzgurg-Landau like double-well term:
\begin{equation}
\label{eq:fgl}
H_{1}=\int d^d{x} \left(a e_1 -\frac{b}2 e_1^2 
+ \frac{c}4 e_1^4 + \frac{\kappa}2 |\nabla e_1|^2 \right),
\end{equation}
The remaining $d(d+1)/2-1$ independent components of the strain tensor
give the harmonic contribution: 
\begin{equation}
  \label{eq:fh}
H_h= \mu \int d^d{x} \sum_{i,j=1}^{d}\widetilde{\epsilon}_{ij}\,^2.
\end{equation}
where $\mu$ is the shear modulus of 
the material, and the (traceless) `accommodation strains' 
tensor $\widetilde{\epsilon}_{ij}\equiv\epsilon_{ij}
-\delta_{ij}e_1/\sqrt{d}$ has been defined.
\begin{figure}[tbp]
\includegraphics[width=8cm,trim=0 180 0 70,clip=true]{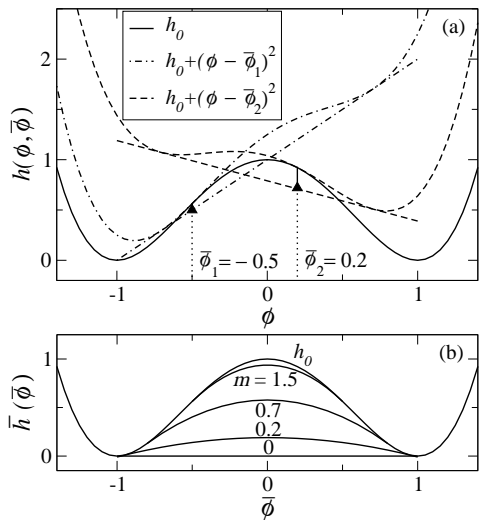}
\caption{(a) $\bar\phi$-dependent common tangent construction. The cases 
$\bar\phi=-0.5$ and $0.2$ are illustrated ($m=1$). The triangles indicate the
minimum value of $h$ attainable for the corresponding values of $\bar \phi$. 
(b) Analytical results for $\bar h(\bar\phi)$ as a function 
of $\bar\phi$ for $\gamma=0$ and
different values of $m$, as indicated. For $m>2$ the solution is 
$\bar h(\bar\phi) = h_0(\bar\phi)$.}
\label{fig:fdf}
\end{figure}

\paragraph{Zero temperature theory }
In the absence of dislocations and in order to be physical
the strain tensor must fulfill the St. Venant compatibility 
condition\cite{cha94}: 
\begin{equation}
  \label{eq:com}
{\rm curl}\; {\rm curl}\; \epsilon={\bf 0}
\end{equation}
which leads to constraints between $e_1$ and $\widetilde{\epsilon}_{ij}$.
A uniform strain satisfies Eq.~(\ref{eq:com}) trivially and therefore it 
is unconstrained. 
 Thus the ${\bf k}={\bf 0}$ term
in the Fourier decomposition of  $H_h$, corresponding 
to {\em uniform} accommodation strains, 
can be minimized by choosing $\widetilde{\epsilon}_{ij}({\bf k}={\bf 0})=0$.
 For ${\bf k}\ne{\bf 0}$ the accommodation strains
can be eliminated in favor of $e_1$ by minimizing 
$H_h$ with respect to $\widetilde{\epsilon}_{ij}$, and enforcing the 
constraints in Eq.~(\ref{eq:com}) 
by Lagrange multipliers\cite{kar95ras01}. 
The resulting 
equilibrium equations together with Eq.~(\ref{eq:com})
allow Eq.~(\ref{eq:fh}) to be written as a function of 
$e_1$ alone:
$$
H_h=(d-1)\mu\frac1{L^d}\sum_{{\bf k}\ne{\bf 0}} e_1({\bf k})e_1(-{\bf k}).
$$
Here $L$ is the linear dimension of the system. 
A crucial point is that the ${\bf k}={\bf 0}$ term is
excluded from the sum. We can extend the sum to all 
${\bf k}$ by explicitly subtracting the missing term. 
We obtain in real space:
\begin{eqnarray}
  \label{eq:fhde1}
&&  H_h/[(d-1)\mu]= \int d^d{x} [e_1({\bf x})-\bar e_1]^2 \\
&&=\int d^d{x}\int d^d{x}'  
\left[\delta({\bf x}-{\bf x}')-\frac1{L^d}\right] 
e_1({\bf x})e_1({\bf x}')\nonumber
\end{eqnarray}
with $\bar e_1\equiv\int d^d{\bf x} e_1 /L^d$.
The accommodation strains
induce a local term that increases the stiffness at all wave
vectors, and a distance independent ``ferro'' like interaction in
$e_1$ that decreases the uniform stiffness alone,  
i.e., the stiffness at finite 
wave vector is larger than at zero wave vector, a well known result
that holds also for linear elastic solids\cite{cha95}. 
A similar result has been obtained in $d=2$ 
in the context of the dislocation mediated melting 
problem close to an isostructural critical point\cite{cho96}.

We define $\phi\equiv e_1 \sqrt{c/b}$, $m\equiv 4(d-1)\mu/b$ and 
$\gamma\equiv 2\kappa/b$. Up to an additive constant the coarse grained energy 
density $h$ [from Eqs.~(\ref{eq:fgl}) and (\ref{eq:fhde1})]
in units of the height of the double well barrier [$b^2/(4c)$] reads:
\begin{equation}
\label{eq:gamma}
h(\phi,\bar \phi)= h_0(\phi)+\gamma |\nabla \phi|^2 + m (\phi-\bar \phi)^2
\end{equation}
with $h_0(\phi)\equiv (1-\phi^2)^2$ being the double well in the dimensionless
volume $\phi$. The term linear in the OP
has been eliminated by redefining the origin of pressure. 
The parameter $\gamma$ 
fixes the width of the interface
between the low and high volume phases in a mixed state. 

We first discuss the $\gamma = 0$ case and generalize for 
 a finite $\gamma$ below. The energy density 
Eq.~(\ref{eq:gamma}) couples the values of $\phi$ at different 
spatial positions only through $\bar\phi$, namely it corresponds
effectively to a MF. We stress however, that this is the {\em exact}
energy of the model.  

Mixed phase solutions for a constrained total volume  
obey the familiar common tangent construction:
$\partial h(\phi,\bar \phi)/\partial \phi|_{\phi=\phi_-}=
\partial h(\phi,\bar \phi)/\partial \phi|_{\phi=\phi_+}=
[h(\phi_-,\bar \phi)-h(\phi_+,\bar \phi)]/(\phi_- -\phi_+)$
where  $\phi_{\pm}=\pm\sqrt{(1-m/2)}$ and the $+$ ($-$) corresponds 
to the OP of the expanded (collapsed) phase.
  The only difference with a conventional phase separation computation
is that the construction is $\bar \phi$-dependent because 
$h$ depends on the global variable $\bar \phi$ [see Fig.~\ref{fig:fdf}(a)]. 
This dependence makes the total energy of the mixture 
non additive.  In Fig.~\ref{fig:fdf}(b) we report the resulting
average energy density $\bar h(\bar \phi)$.
For $m>2$ the mixed solution does not exist and the system remains 
always uniform. We have checked these results with direct analytical
 computations assuming specific shapes for the coexisting phases
and different boundary conditions.  
In all cases the maximum shear strain is of the same order as the 
relative volume mismatch. As long as the volume mismatch is small 
(a few percent)
the shear strain will behave elastically. For large volume mismatch
glide of atomic planes may take place and our theory becomes
approximate. In any case the transition will be irreversible. 

For a pressure controlled experiment the physical 
pressure is given by $P=p \sqrt{b^3/(16cd)} -a/\sqrt{d}$ with $p$
the dimensionless pressure: $p=-\partial \bar h(\bar \phi)/\partial \bar\phi$ 
at $T=0$. In the mixed phase the free energy has negative curvature
so the compressibility is negative [Fig.~\ref{fig:fdf}(b)].
This thermodynamic state is stable at constant volume with the 
pined boundary conditions of Ref.~\cite{ber76}. This 
violation of usual stability criteria is common in non-extensive 
systems\cite{veg02}.

\begin{figure}[bp]
\includegraphics[width=7cm,trim=0 320 0 60,clip=true]{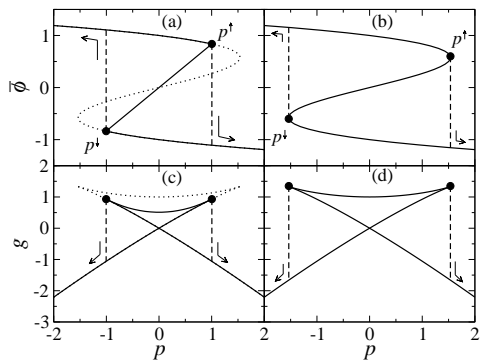}
\caption{Mean volume $\bar\phi$ and Gibbs free energy density 
for $m=0.6$ (left) and $m>2$ (right). Dotted lines are the uniform solution.
Continuous lines are the solutions that minimize 
Eq.~(\protect\ref{eq:gamma}) for each value of $\bar\phi$.
The arrows and dashed lines 
indicate the hysteresis cycle in pressure controlled
experiments. Full dots correspond to $p^{\uparrow}$, $p^{\downarrow}$.}
\label{fig:pdf}
\end{figure}

In Fig.~\ref{fig:pdf} we show the values of
$\bar\phi$ and the Gibbs free energy density $g(p) \equiv \bar h[\bar
\phi(p)]+p\bar\phi(p)$ as a function of $p$ for two values of $m$. 
Controlling the total 
volume the system would follow the continuous line. The straight
portion of this solution in Fig.~\ref{fig:pdf}(a) 
corresponds to the mixed phase. 
Consider now a pressure controlled experiment for $m<4/3$ 
[Fig.~\ref{fig:pdf}(a) and (c)].  
Starting from the high volume phase at low pressure and increasing 
$p$ the system remains uniform until $p^{\uparrow}$, given by
 the point in which the mixed solution line
intersects the uniform solution. At this point the small volume phase
nucleates spontaneously since $\gamma = 0$.
It is easy to show that the mixed state is
mechanically unstable at constant pressure due to the negative 
compressibility, so the volume collapses and the 
system jumps to the lower branch of the curve following the vertical
line.  Decreasing  the pressure the process is reverted and the volume
suddenly expands at $p^{\downarrow}(=-p^{\uparrow})$ producing the 
hysteresis loop indicated in the figure. A conventional first-order
phase transition driven by pressure at a very low rate 
(so that nucleation occurs) would follow 
a vertical line at $p=0$ in the top panel. 
This is not possible here because the infinite range 
interaction makes the critical drop radius\cite{cha95} diverge, and
the nucleation energy-barrier to scale as the
volume of the system.

In the case $m>4/3$ 
the compressibility of the uniform solution diverges at the spinodal
points [full dots in Fig.~\ref{fig:pdf} (b) and (d)].
The transition proceeds by a ${\bf k}={\bf 0}$ instability 
at $p^{\uparrow}=-p^{\downarrow}=8/(3\sqrt{3})$, 
independent of $m$ and with no energy barrier. 
The lower panels in Fig.~\ref{fig:pdf} show that the
transition involves an unavoidable jump in the thermodynamic potential 
for all  $m\ne 0$ justifying calling this transition `zeroth 
order'\cite{veg02}.

\begin{figure}[bp]
\includegraphics[width=7.5cm,trim=0 440 0 50,clip=true]{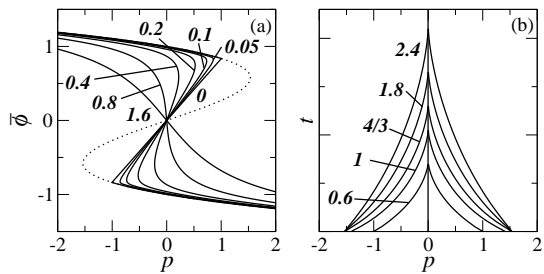}
\caption{(a) $\bar \phi$ vs. $p$ for $m=0.6$ and different values
of the dimensionless temperature $t$, as indicated. 
(b) $t-p$ phase diagram for $\gamma=0$ and 
various $m$ values (indicated in the figure). The interior of the wedge
(for each value of $m$) is a region in which nucleation is forbidden.
}
\label{fig:tem}
\end{figure}

\paragraph{Finite temperature theory }
Using MF theory it is straightforward to obtain 
the exact partition function for $\gamma=0$, and 
the exact $\bar\phi-p$ dependence at different temperatures.
In Fig.~\ref{fig:tem} we show
the isotherms in the $\bar \phi-p$ plane for $m=0.6$ for different 
values of a dimensionless temperature $t$ (units are restored below). 
The pressure width ($\Delta p$) and volume width ($\Delta \bar\phi$) of the 
hysteresis loop is reduced at finite temperatures 
up to the point where it vanishes, which determines the
critical point coordinates $(p_c=0,t_c(m))$. The  pressure
width as a function of temperature determines a wedge of forbidden nucleation
[shown in Fig.~\ref{fig:tem}(b) for different values of $m$] 
around the critical pressure $p_c=0$. 

For $m<4/3$ the singularities at $p=\pm p^{\uparrow}$ 
of the $T=0$ isotherm become rounded at finite temperature  
[Fig.~\ref{fig:tem}(a)] due to the thermally induced spatial fluctuations
of the OP. This implies that the compressibility will show precursor
effects for all $m>0$, diverging at the transition.

The solution in the case of finite $\gamma$ and $m$ 
is given implicitly by  
$\bar\phi=\bar\phi_0(p+m\bar\phi)$ where 
$\bar\phi_0(p)$ is the equation of state of 
the same model except that $m=0$, i.e.,
the well studied $\phi^4$-model which has the 
universality class of the short-range Ising model\cite{cha95}. 
If $m>4/3$ the qualitative description of the forbidden nucleation 
region are completely equivalent to the $\gamma=0$ case. For $m<4/3$ a region 
of normal nucleation and growth appears around the wedge of forbidden 
nucleation at low temperatures and the divergence of the
compressibility 
gets a cutoff. These effects never reach the critical point. Both for $\gamma=0$
and $\gamma>0$ we find that critical exponents are given exactly by MF. 
This is also in agreement with the $d=2$ theoretical analysis of 
Ref.~\cite{cho96}.

\paragraph{Comparison with experiments}
A wedge where nucleation does not occur is 
observed in Ce\cite{gsc79,law81}, SmS\cite{ras78} and related
 systems. Interestingly it is also observed in a
volume-collapse transition in amorphous ice\cite{mis94}.
However without detailed studies it is difficult 
to be sure that it does not arise from usual hysteresis
due to a finite driving rate, as occurring in other first order transitions. 
Other indicators however, point more strongly to the applicability
of our theory.

In agreement with the present theory MF  critical exponents have been
 observed\cite{law81}.
The connection between MF critical behavior and long-range strains  
has long been suspected\cite{law81}.

In contrast with usual first-order phase transitions, where 
hysteresis is not intrinsic, one can define here a critical exponent for 
the pressure width of the hysteresis loop: $\Delta p\propto |t-t_c|^{3/2}$. We are not
aware of detailed measures of this critical exponent, however
the curvature  of the boundaries  found in Ref.~\cite{ras78} for 
Gd doped SmS are compatible with this exponent. 
We hope our work will stimulate further experimental work.

Also in contrast with usual first-order transitions, critical behavior 
(although classic like) can be easily observed  far from the critical point.
The divergence of the compressibility discussed above
implies that the bulk modulus vanishes as $|p-p^{\uparrow}|^{1/2}$ and 
jumps discontinuously to a finite value for $p>p^{\uparrow}$.
 Decreasing pressure the jump and the square root singularity change 
side and occur at $p^{\downarrow}$.
Similar behavior is found as a function of temperature.

An anomaly 
measured in YbInCu$_4$ on cooling resembles our prediction when plotted in 
an appropriate scale (see inset of Fig.~4 in Ref.~\cite{bkin94}). 
This also 
explains the appearance of precursor effects which have been very puzzling 
in the mixed valence literature and have led some authors to interpret the 
transition as second  order\cite{bkin94}.

In mixed valence systems the double well is attributed to the 
anomalous contribution of $f$-electrons to the crystal 
binding\cite{all82joh95}.
For a  realistic description it may be necessary to take the parameters in
Eq.~(\ref{eq:fgl}) to depend on temperature to take into account  
temperature effects of the  $f$-electrons. In fact a trivial linear 
and quadratic temperature dependence of $a$ fits our symmetric 
phase diagram to the asymmetric forms of Ce\cite{gsc79,law81} 
and SmS\cite{law81,ras78}. 

The temperature dependence of $a$ does not affect
 the critical temperature, but $b$ does.
Indeed in current explanations of the critical  point in $f$-electron
systems this electronic effect is crucial whereas the shear 
rigidity does not play a relevant role\cite{all82joh95}. 
To check the relevance of the temperature scale set by the shear modulus
on determining real values of $T_c$ we 
take the opposite point of view, i.e., we neglect the temperature 
dependence of the coefficients in  Eq.~(\ref{eq:fgl}) and take into
account the effect of the shear rigidity alone. 
Restoring dimensions, the critical temperature is given by:
$k_B T_c=t_c(m) v_0  b^2/(4c)$  where $v_0$  is
the minimum volume in the system allowed to fluctuate independently.
For the case $\gamma=0$ this volume should be assimilated
to the atomic volume. In the case $\gamma>0$, this volume 
can be greater depending on the strength of the short distance interactions 
parametrized by $\gamma$. In any case we expect it to be of the 
order of a few atomic volumes.  
The parameters $b$, $c$ and $m$ can be fixed by fitting
the observed pressure and volume with of the hysteresis loop at 
a given temperature and from
the measured value of the shear modulus\cite{gsc79,hai84}. 
Within the experimental uncertainties this procedure gives the 
right value of $T_c$ for Ce ($\sim 550$K\cite{gsc79,law81}) and 
SmS  ($\sim 750$K\cite{law81,ras78}) if we assume $v_0\sim 2 v_f$ where 
$v_f$ is the volume associated with one formula unit.   
Since the relation between $v_0$ and $v_f$ go beyond our model 
we cannot exclude  temperature effects on the coefficients of 
Eq.~(\ref{eq:fgl}) contributing to $T_c$\cite{all82joh95} 
however the fact that we obtain the right 
order of  magnitude with a  $v_0$  on the order of $v_f$ implies that the
scale set by the shear modulus can not be neglected in the determination of 
 $T_c$. 

In conclusion 
we have presented an exactly solvable model of an isostructural
volume collapse transition, the solid-state analog of the  van der Waals
 transition. Unlike the fluid counterpart the thermodynamics 
is very anomalous due to the presence of infinite range interactions. 
In particular, the transition is intrinsically irreversible 
with a discontinuity of the thermodynamic potential making of it a  
laboratory example of a class of phase transitions, namely zero order
transitions, first discussed in the context of gravitational 
systems\cite{veg02}.

Despite the simplicity of our model, which assumes an isotropic solid,
we believe the present features are quite robust. Indeed our findings are 
in good agreement with the behavior of real anisotropic 
systems\cite{gsc79,law81,ras78,bkin94}. In these mixed valence systems 
the unexpected thermodynamics is a consequence of the interplay between 
strong correlations, determining the double well potential\cite{all82joh95}, 
and long-range strains. The present approach can be seen as a first step to 
understand this interplay in more complex non-isostructural transitions 
like the magnetoresistant manganites\cite{lor01Ilor02} which also show volume
mismatch among different phases and irreversibility\cite{cox98}.

\end{document}